\documentclass[aps,prl,twocolumn,groupedaddress,floatfix]{revtex4}

\usepackage{times}
\usepackage{color}
\usepackage[dvips]{graphicx}
\usepackage{amsmath}
\usepackage{amsfonts}
\usepackage{amssymb}

\begin{document}

\title{Kinetically driven ordered phase formation in binary colloidal crystals}

\author{D. Bochicchio$^{a}$, A. Videcoq$^{b}$, R. Ferrando$^{a}$\footnote{Corresponding author, e-mail ferrando@fisica.unige.it}}

\affiliation
{$^{a}$ Dipartimento di Fisica and CNR-IMEM, Via Dodecaneso 33, Genova, I16146, Italy,
$^{b}$ SPCTS, UMR 7315, ENSCI, CNRS; Centre Europe\'een de la C\'eramique, 12 rue Atlantis, 87068 Limoges cedex, France
}



\begin{abstract}
The aggregation of binary colloids of same size and balanced charges is studied by Brownian dynamics simulations for dilute suspensions. It is shown that, under appropriate conditions, the formation of colloidal crystals is dominated by kinetic effects leading to the growth of well-ordered crystallites of the sodium-chloride (NaCl) bulk phase.  These crystallites form with very high probability even when the cesium-chloride (CsCl) phase is more stable thermodynamically. Global optimization searches show that this result is not related to the most favorable structures of small clusters, that are either amorphous or of CsCl structure. The formation of the NaCl phase is related to the specific kinetics of the crystallization process, which takes place by a two-step mechanism. In this mechanism, dense fluid aggregates form at first and then crystallization follows. It is shown that the type of short-range order in these dense fluid aggregates determines which phase is finally formed in the crystallites. The role 
of hydrodynamic effects in the aggregation process is analyzed by Stochastic Rotation Dynamics - Molecular Dynamics simulations, finding that these effects do not play a major role in the formation of the crystallites.
\end{abstract}

\maketitle

\section{Introduction}
Colloidal suspensions have been widely studied for a variety of applications. From the point of view of basic science, a growing interest stems from the fact that colloidal suspensions are many-particle systems that are easier to observe than molecular systems. Aggregation and nucleation phenomena of colloids are therefore simpler to analyze, also by means of real-space techniques. Moreover, the interactions between colloids can be tuned in a controlled way so that a variety of structures with different properties can be experimentally produced \cite{Leunissen2005nature}. Recently, much attention has been devoted to binary colloids in which two types of colloidal particles (types A and B in the following) can acquire opposite charges in suspension, interacting mainly through screened electrostatic forces. Binary colloids can thus  aggregate \cite{Lopez2006softmatter,Rollie2008langmuir,Lebovka2012note}, in a process which is usually called \textit{heteroaggregation}. The competition between 
attractive and repulsive forces gives rise to a rich phenomenology, which is the subject of active investigation. In fact, binary colloids have been shown to be able to  form either compact colloidal crystals with a good degree of long-range order \cite{Leunissen2005nature,Bartlett2005prl,Hynninen2006prl}, or colloidal gels characterized by a network of fractal-like aggregates \cite{Sanz2008jpcm,Piechowiak2012pccp,Russell2012softmatter}, the latter being more tenuous than in single-component gels \cite{Russell2012softmatter}.

From the theoretical/computational point of view, notable attention has been devoted both to the determination of the equilibrium properties of binary suspensions \cite{Lopez2006softmatter,Sanz2008jpcm,Bier2010jcp,Pavaskar2012jcp}, with the calculation of phase diagrams of representative model systems, and to their aggregation kinetics \cite{Lopez2006softmatter,Sanz2007prl,Sanz2008jpcm,Cerbelaud2010softmatter,Piechowiak2010langmuir}. However, several aspects of heteroaggregation are still to be understood, especially for what concerns the basic mechanisms that determine the structures obtained in the actual aggregation processes.

An important issue in this colloidal aggregation is to determine to what extent the structures that are actually formed in the growth of colloidal crystals reflect their thermodynamic equilibrium structures. This has been investigated in the case of the heteroaggregation of colloids interacting through screened electrostatic forces by Sanz et al. \cite{Sanz2007prl}, in the case of high volume fraction $\Phi \simeq 0.47$. In their simulations, Sanz et al. used the forward-flux sampling method \cite{Allen2005prl} to determine nucleation rates. They found that highest nucleation rate was associated to the formation of a substitutionally disordered fcc crystal, which is neither the lowest free-energy phase nor the phase characterized by the lowest free-energy barrier for nucleation from the melt. In fact, the lowest free-energy structure is of cesium-chloride (CsCl) type. Sanz et al. attributed this finding to the fact that sub-critical small clusters have disordered fcc structures, so that growth continues 
keeping this kind of structure because the transition to the CsCl structure is kinetically inhibited.    

In this article we perform Brownian dynamics simulations of crystal nucleation and growth in the same model system as Sanz et. al. \cite{Sanz2007prl,Sanz2008jpcm}, consisting of binary spherical colloids of same radius $a$ and absolute value of the charge, interacting through a screened electrostatic potential of Yukawa-like form. This potential gives a valid description of the interactions between colloids in index-matched solvents \cite{Leunissen2005nature,Bier2010jcp}. For a given colloid volume fraction $\Phi$, the equilibrium behavior of colloids interacting by this potential can be described as a function of two parameters: $U^*=U_0/(k_B T)$, where $U_0$ is the well depth, and $\kappa a$, where $\kappa^{-1}$ is the Debye screening length. At variance with Refs. \cite{Leunissen2005nature,Sanz2007prl,Sanz2008jpcm},  we consider a quite dilute system, with  a volume fraction $\Phi=0.05$, which is below the percolation threshold. 

Our simulations show that aggregates corresponding to metastable colloidal crystals form with very high probability under appropriate conditions. 
While in the case of high volume fractions treated in Ref. \cite{Sanz2007prl} the metastable phase was disordered, in our dilute system we find that a very well ordered metastable phase form.
Specifically, we show that crystallites with the sodium-chloride (NaCl) structure grow for values of $(U^*,\kappa a)$ at which the cesium-chloride (CsCl) crystal structure is the most stable from the thermodynamic point of view.  The formation of the NaCl-type crystallites is rationalized in the framework of the two-step mechanism of crystal nucleation \cite{Lutsko2006prl,Vekilov2010nanoscale,Schilling2010prl}, in which a dense colloidal fluid acts as a precursor for crystal nucleation. We show specifically that the type of short-range order in this dense fluid is crucial for determining which crystal phase is going to be formed. We show also that the NaCl phase formation is not related to energetically favorable structures of small clusters, the latter being either amorphous or with CsCl structure. Colloidal crystals with the NaCl structure were obtained with colloids of different sizes \cite{Vermolen2009pnas}. Our results demonstrate these crystals can grow in systems with balanced size and charge. We 
analyze also the importance of hydrodynamic effects in the nucleation of the NaCl phase by performing Stochastic Rotation Dynamics - Molecular Dynamics (SRD-MD) simulations \cite{Sh:1,Sh:2,Sh:3}, which show that these effects do not play a major role.


\section{Model and Methods}
In our simulations the interaction potential between the colloids is of the form
\begin{equation}
 U_{ij}(r) = U_0 \, \mathrm{sign} (q_i q_j)  \frac{2a}{r} \mathrm{e}^{-\kappa (r - 2a)}
\end{equation}
where $U_0$ is a constant, which gives the value of the potential at contact, and the sign of the charge $q_i$ depends on the colloid $i$ and being either of type A or B. In the following we use $U^*= U_0/(k_BT) =9$. This expression of the potential is valid for $r \geq 2a$. For $r < 2a$, the hard-wall repulsion is approximated by a power law increasing as $1 / r^{36}$ \cite{Sanz2008jpcm}.
We have checked that, in infinite crystals, the Madelung energy favors the CsCl phase over the NaCl phase for all values of $\kappa a$, as pointed out in Ref. \cite{Leunissen2005nature}. The Madelung energy difference between the two phases decreases with decreasing $\kappa a$. In Ref. \cite{Leunissen2005nature} it was also shown that, at finite temperature, entropic contributions increase the stability of the NaCl bulk phase, which should however become more stable than the CsCl phase only below $\kappa a \simeq 2.5$ for $U^*$ of the magnitude considered here.  

In the following, we use three different simulation techniques, Brownian Dynamics (BD) simulations, Basin Hopping (BH) global optimization searches, and SRD-MD simulations.

Our BD simulations are performed by numerically solving the Langevin equation by means of the algorithm described in Ref. \cite{Piechowiak2012pccp}. We choose the dimensionless time unit as  $t^*= 4 a^2/D_0$, where $D_0$ is the diffusion coefficient of isolated colloids in the suspension. $t^*$ is the time taken by an isolated colloid to diffuse across a distance equivalent to its diameter. 
$D_0$ is given by Einstein's relation, in which the solvent friction and the colloid mass are chosen as in Ref. \cite{Piechowiak2012pccp}.
The Langevin equation is solved by using time steps $\delta t$  in the range  $10^{-5} - 10^{-6} \, t^*$. The number of colloids $N$ in the simulations is 200, half being positively and half negatively charged. The colloids are initially placed in a cubic box of volume $V$ with random positions. Periodic boundary conditions are applied. Simulations are stopped after a time $t_{fin} \simeq 2 \, 10^3 \, t^*$, which is sufficient to let all colloids form a single aggregate of crystalline structure. For comparison, in the 
system of Ref. \cite{Piechowiak2012pccp}, this $t_{fin}$ would amount to $\sim 10^3$ s.

The lowest-energy structures of colloidal clusters have been searched for by a global optimization procedure employing the BH method \cite{Wales1997}, which has proven to be very efficient in the global optimization of atomic clusters. This method consists of a Metropolis Monte Carlo procedure in which the energy is locally minimized at each step. BH has been applied recently also to colloidal clusters in Ref. \cite{Cerbelaud2010jcp}, where the computational procedure is described (see also \cite{Rossi09jpcm} for a more complete description of the method). We note here that we have used three different kinds of elementary moves to generate new configurations in the Monte Carlo procedure. These are the \textit{Brownian} move, the \textit{shake} move \cite{Rossi09jpcm}, and a combined move, made of an uniform compression, in which the whole cluster is compressed by a factor 0.8 - 0.9, followed by a mild shake, i.e. of small random displacements of all colloids around their positions.
 After each move, the cluster is locally relaxed to generate the new configuration, which is then either accepted or refused according to the Metropolis rule. In all cases, 5 simulations of 10$^5$ steps each have been performed. 

Hydrodynamic effects have been taken into account by means of SRD-MD simulations.
SRD-MD is a combination of Stochastic Rotation Dynamics and standard Molecular Dynamics
that allows taking into account long range hydrodynamic effects in the study of colloidal
suspensions \cite{Sh:1,Sh:2}. This technique has been chosen for its relative efficiency and low computational
cost, to check whether hydrodynamic effects could affect the nucleation process in our system. 
SRD-MD is based on the use of fake fluid particles that mimic the fluid properties
(essentially its momentum). The simulation box is divided into smaller cubic cells. In each cell the
fluid particles exchange momentum thanks to a rotation of their velocities.
For an accurate description of the method we refer to the literature \cite{Sh:1,Sh:2,Sh:3}. Here we report only the choices of the most
important parameters for our system. The linear size of the cells $a_0$ is taken as half of the radius of the colloids. For
the key parameters of the SRD-MD simulations we made the same choice proposed by Padding and Louis \cite{Sh:2}:
the number of fluid particles per cell $\gamma$ equal to 5, the dimensionless mean free path $\lambda$ equal to 0.1
and the rotation angle $\alpha$ equal to 90$^o$. The parameters have been chosen to reproduce the same diffusion coefficient of isolated colloids in both BD and SRD-MD \cite{Sh:3}. Recently it has been proved that these choices are able to
correctly reproduce the essential features of the hydrodynamic interactions in a colloidal
suspension \cite{Sh:3}. 
For what concerns the energetic model, we have chosen the same Yukawa potential used in BD simulations for
colloid-colloid interactions, while for colloid-fluid interactions we have chosen a soft-sphere potential
\begin{eqnarray}
 V_{cf} (r) &  = &  \varepsilon_{cf} \left ( \frac{r}{\sigma_{cf}} \right )^n 
\end{eqnarray}
for $r \leq r_c = 2.5 \, \sigma_{cf}$ and 0 otherwise.
We have chosen $n$=12 and $\varepsilon_{cf}= 2.5 \, k_B T$. The value of $\sigma_{cf}$ has been taken smaller than the colloidal radius ($\sigma_{cf}=0.8 \, a$) to avoid spurious depletion attractions between colloids \cite{Sh:2}.

\section{Results and Discussion}

\begin{figure}[ht]
\includegraphics[width=8.5cm]{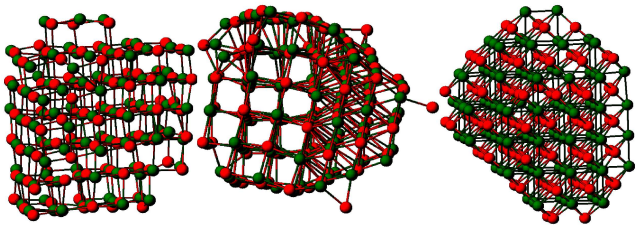}
\caption{\label{fig1}. From left to right, aggregates of NaCl, mixed and CsCl structures. Particles of different charges are shown in green (dark grey) and red (light grey).}
\end{figure}

Let us consider the aggregation of the colloids for $\kappa a=3$, at which the NaCl crystal phase is never found to be the most stable \cite{Leunissen2005nature,Bier2010jcp} from the thermodynamic point of view. The results concerning final aggregates in BD simulations (see Table \ref{table1}) show that in a large majority of cases (15 in 20 simulations) structures of NaCl type are formed. The remaining simulations give either mixed or CsCl structures, the latter being found only in two cases. Representative snapshots of NaCl, mixed and CsCl structures are shown in Fig. \ref{fig1}.  Also the simulations for $\Phi=0.08$ give a large majority of NaCl-type aggregates. As $ \kappa a$ increases  the CsCl phase becomes more and more energetically favorable. Therefore, we expect that there should be a critical value $(\kappa a)_c$ at which the probability 
of 
forming both types of aggregates is the same. From the results in Table \ref{table1}, it turns out that  $(\kappa a)_c$ is close to 3.3, a value which is significantly larger than the $\kappa a$ at which the bulk NaCl and CsCl phases coexist in thermodynamic equilibrium \cite{Leunissen2005nature}. Increasing $\kappa a$ further leads to the formation of CsCl-type aggregates with an overwhelming probability. On the other hand, if $\kappa a$ is decreased to 2.55, NaCl-type crystallites are always formed.

\begin{table} [h]
 \begin{tabular}{llccc}
 $\Phi$ \; \;    &  $\kappa a$ \; \;   & \; NaCl \; &  \; mixed \;   & \; CsCl \;  \\
\hline
 5\%   &  2.55    &  10 & 0   & 0    \\
 5\%   &  3.00    &  15 & 3   & 2    \\   
 8\%   &  3.00    &   8 & 1   & 1   \\   
 5\%   &  3.30    &  10 & 4   & 6    \\ 
 5\%   &  3.75    &   1 & 0   & 9   \\ 
 5\%   &  5.00    &   0 & 0   & 10   \\ 
 5\%   &  9.00    &   0 & 0   & 10   \\   
\hline
\end{tabular}
\caption{\label{table1} Final structure of the aggregates obtained in Brownian simulations for different values of $\kappa a$ and volume fraction $\Phi$. Either 10 or 20 simulations have been performed in each case.}
\end{table}

What are the driving forces leading to the formation of these aggregates of the metastable NaCl bulk phase? 

\begin{figure}[ht]
\includegraphics[width=8.5cm]{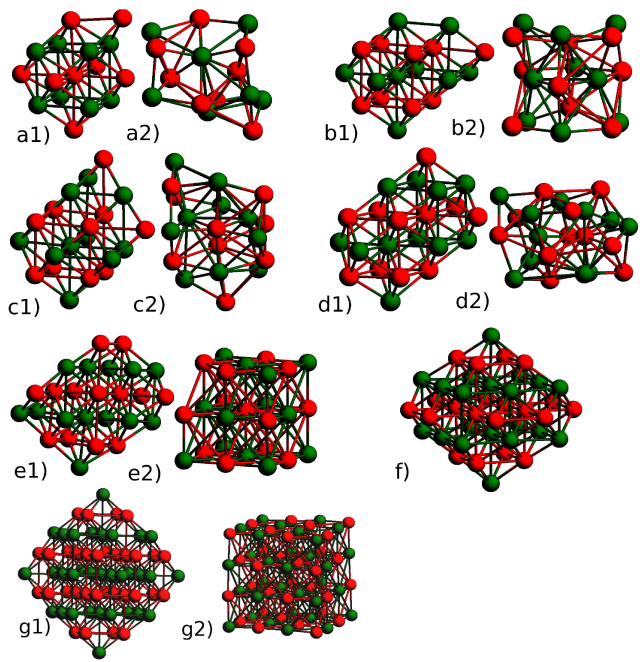}
\caption{\label{fig2} Lowest-energy structures of the aggregates as determined by global optimiziation searches, in the cases of A$_{8}$B$_{8}$ (a1 and a2), A$_{9}$B$_{9}$ (b1 and b2), A$_{10}$B$_{10}$ (c1 and c2), A$_{11}$B$_{11}$ (d1 and d2), A$_{13}$B$_{13}$ (e1 and e2), A$_{20}$B$_{20}$ (f) and A$_{32}$B$_{32}$ (g1 and g2). The actual lowest-energy structure depends on $\kappa a$ as explained in the text.}
\end{figure}

A possible explanation of this result may originate from the fact that the aggregation process starts with the formation of very small colloidal clusters that subsequently grow and coalesce, as shown in the simulation of the aggregation in a similar system \cite{Piechowiak2012pccp}. Therefore, the structure of the final aggregate may preserve memory of the equilibrium structures of small aggregates. In order to investigate this point we have searched for the lowest-energy structures of colloidal clusters in the size range from A$_{8}$B$_{8}$ to A$_{32}$B$_{32}$. The main structural motifs singled out for the $\kappa a$ values reported in Table \ref{table1} are shown in Fig. \ref{fig2}, where  a few selected sizes are considered (A$_{8}$B$_{8}$, A$_{9}$B$_{9}$, 
A$_{10}$B$_{10}$, A$_{11}$B$_{11}$, A$_{13}$B$_{13}$, A$_{20}$B$_{20}$, and A$_{32}$B$_{32}$). Depending on $\kappa a$, the dominant motif may change as follows.

For small aggregates, from A$_{8}$B$_{8}$ to A$_{10}$B$_{10}$, the lowest-energy structures  either belong to the CsCl motif, as structures a1, b1, and c1 in Fig. \ref{fig2}, or are disordered (i.e. they are neither of CsCl- nor of NaCl-type, see a2, b2, c2 in Fig. \ref{fig2}). CsCl-type structures prevail at $\kappa a \geq 5$, while at smaller $\kappa a$ disordered structures are more favourable.  From A$_{11}$B$_{11}$ on, CsCl-type structures (d1, e1, f, g1 in Fig. \ref{fig2}) always prevail  with a single exception which is found at $\kappa a=2.55$ only. For this $\kappa a$, the best structure of A$_{13}$B$_{13}$ is of NaCl type, being a 3$\times$3 cube with one vertex missing (structure e2). However, with increasing cluster size NaCl structures become more and more unfavorable. In fact, at the next \textit{magic} size for NaCl structures, i.e. at size 64, where a perfect NaCl cube  A$_{32}$B$_{32}$ can be formed (structure g1), the CsCl structure is lower in energy for all $\kappa a$ down to 2.55.

These results show that structures of CsCl type prevail already for small aggregates, indicating that there are no evident effects that favor the NaCl structures at small sizes with respect to what is found in bulk crystals. Therefore the explanation of the growth of NaCl-type aggregates  originates from a different effect, which is better understood by analyzing in details the aggregation process in the simulations.

To this end, we define two order parameters that are able to discriminate between the different crystalline phases. These parameters (referred to as $P_1$ and $P_2$ in the following) are defined as follows.  $P_1$ is the fraction of colloids having more than 6 first neighbors. This indicates the possible presence of CsCl-type short range order, because in NaCl-type aggregates the number of first neighbors of a given colloid cannot be larger than 6. On the other hand, $P_2$ is related to the configuration of second neighbors, being defined as 
 \begin{equation}
  P_2 = \frac{1}{N} \sum_{i \neq j}  \left[ \frac{1}{12} \mathrm{e}^{- \frac{(r_{ij}-r_{2N})^2}{2 \sigma^2}}- \frac{1}{6} \mathrm{e}^{- \frac{(r_{ij}-r_{2C})^2}{2 \sigma^2}}\right],
 \end{equation}
where $r_{ij}$ is the distance between colloids $i$ and $j$ ($i$ and $j$ being of the same type), $r_{2N}=2 {\sqrt 2}  a$ and $r_{2C}=4a/{\sqrt 3 }$ are the average equilibrium distance of second neighbors in the NaCl and CsCl structures, and $\sigma$ takes into account the effect of vibrations (in our simulations we choose $\sigma = 0.15 a$, a value that is more than three times smaller than $r_{2N}-r_{2C}$). Positive and negative values of $P_2$ respectively indicate the prevalence of either NaCl- or CsCl-type short-range order in the aggregate. In perfect NaCl and CsCl infinite crystals $P_2$ is close to 1 and -1, respectively.  

\begin{figure}[ht]
\includegraphics[width=8.5cm]{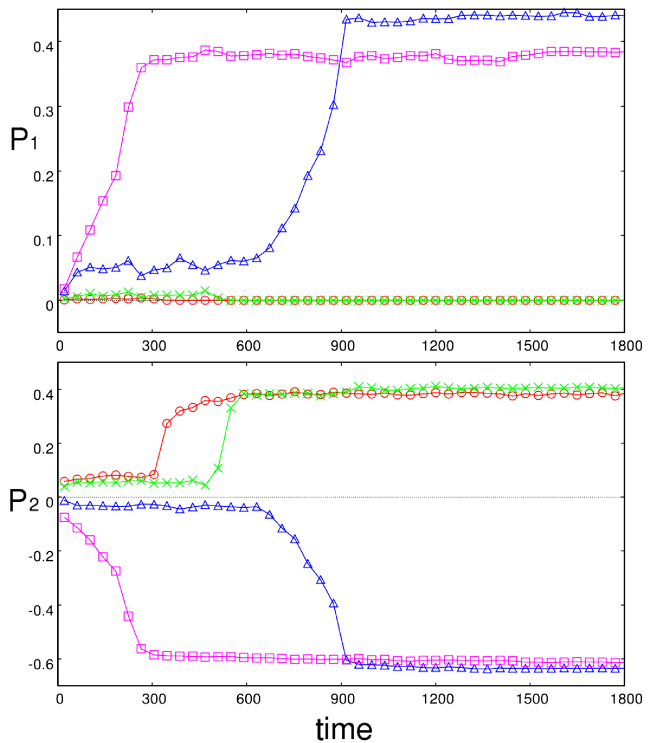}
\caption{\label{fig3} Behavior of order parameters $P_1$ and $P_2$ as functions of time for representative simulations at different $\kappa a$: 2.55 (circles), 3.00 (crosses), 5.00 (triangles) and 9.00 (squares). Time is measured in units of $t^*$. The curves are related to single simulations, but both $P_1$ and $P_2$ have been averaged each 10$^3$ steps in order to give smoother behaviors.}
\end{figure}

\begin{figure}[ht]
\includegraphics[width=8.5cm]{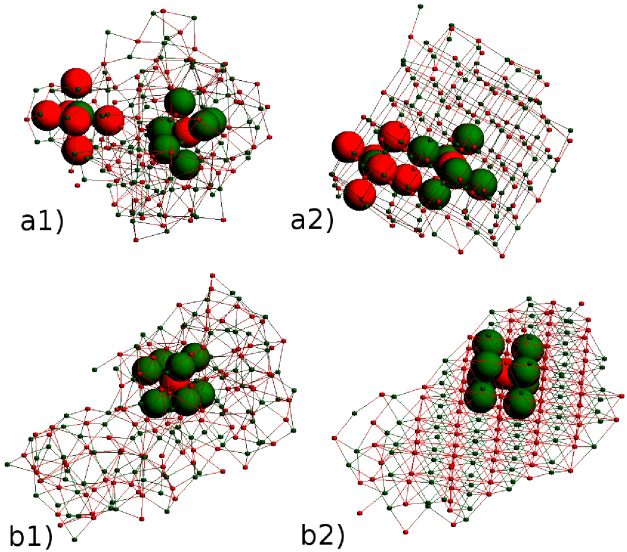}
\caption{\label{fig4} a1) Snapshot from a simulation at $\kappa a =3$ in which the aggregate is still in the disordered state. Two colloids surrounded by a NaCl-type short-range order are shown as big spheres, together with their six nearest neighbors. a2) The final structure in the same simulation, showing a well-ordered  NaCl-type crystallite. b1)  Snapshot from a simulation at $\kappa a =5$ in which the aggregate is still in the disordered state. One colloid surrounded by a CsCl-type short-range order is shown as a big sphere, together with its eight nearest neighbors. b2) The final structure in the same simulation, showing a well-ordered CsCl-type crystallite. }
\end{figure}

Let us consider the typical behavior of a simulation for $\kappa a=3$ (see Fig. \ref{fig3}, crosses). After a rather short time, essentially all colloids attach to some aggregate and, as in most cases, a single aggregate is quickly formed (typically for $t < 100 \, t^*$).  $P_1$ stays very close to 0, showing that the probability of configurations in which a colloid has more than six first neighbours is very low. On the other hand, $P_2$ increases from the initial value of 0 to a positive value $P_2 \simeq 0.05$, indicating that some NaCl-type short-range order is present. However, if we look at the aggregate itself, we find configurations as those reported in the snapshot a1 of Fig. \ref{fig4}. The aggregate does not show any long-range order, but a liquid-like structure, in which some colloids surrounded by a NaCl-type short-range order can be singled out. After a lengthy evolution in this disordered phase, there is a rather sudden increase of $P_2$, which jumps to a value of 0.4, corresponding to the 
transformation of the disordered aggregate into an ordered NaCl-type crystallite, as the one shown in snapshot a2 of Fig. 4.

This behavior can be compared to the case of $\kappa a=5$. Again, a single aggregate is rather quickly formed. The aggregate does not show long-range order, but a somewhat prevailing CsCl-type short-range order (see b1 in Fig. \ref{fig4}) corresponding to a negative $P_2$, and to a small but significantly positive $P_1$. After some time, the sudden transition to the ordered crystallite takes place, but this time the structure is of CsCl type (b2 in Fig. \ref{fig4}).  

This analysis shows that the final structure of the crystallite is therefore well described by the two-step mechanism of crystal nucleation \cite{Lutsko2006prl,Vekilov2010nanoscale}. Starting from diluted configurations ($\Phi \leq 0.08$) dense aggregates form with a metastable liquid-like structure which survives for some time. However, a transition to an ordered structure finally takes place, and the type of ordered structure is correlated, with very high probability, with the features of short-range order in the metastable liquid, as described for example by the parameter $P_2$.  Specifically we find that the final structure is almost always of NaCl type for disordered aggregates in which $P_2 \geq 0.05$, while for $P_2 \leq 0.02$ the CsCl structure is found in most cases. This means that the final ordered structure can be predicted with high probability by inspecting the relative positions of the colloids in the disordered state. We note that this is not always the case in colloid 
crystallization, as demonstrated for hard-sphere glasses in Ref. \cite{Sanz2011prl}.  

The two-step crystallization process in hard-sphere glasses has been studied by Monte Carlo simulations by Schilling et al. \cite{Schilling2010prl}, at high volume fraction $\Phi=0.54$. They found a rather gradual development of crystallinity from the disordered state as the simulation goes on, while in our case crystallization is rather sudden. 

In the disordered phase, for $\kappa a \leq 3.3$, the range of the repulsion between colloids with the same charge is sufficiently long to prevent the  occurrence of configurations in which more than six colloids of the same sign can surround one colloid of the opposite sign. These configurations would be indeed energetically favorable, because the CsCl phase is still the most stable from this point of view, but they are very difficult to reach in an aggregation process starting from a dilute volume fraction. This is the cause of the formation of crystallites of the metastable NaCl phase. Once formed, these crystallites can survive for long lifetimes.

Rather surprisingly, we note that the formation of NaCl-type crystallites is indeed easier when this bulk phase is metastable. We have in fact tried aggregation simulations for $\kappa a=1$ and $U^*=9$, a situation in which the NaCl bulk phase is more stable than the CsCl phase due to entropic effects \cite{Leunissen2005nature}. However, the long range of the repulsion between particles with the same charge renders the aggregation itself quite difficult, so that we were not able to observe any crystallization within $t \simeq 2 \, 10^3 \, t^*$. 

Finally we discuss the importance of hydrodynamic effects by comparing the results of BD and SRD-MD. 

\begin{figure}[ht]
\includegraphics[width=8.5cm]{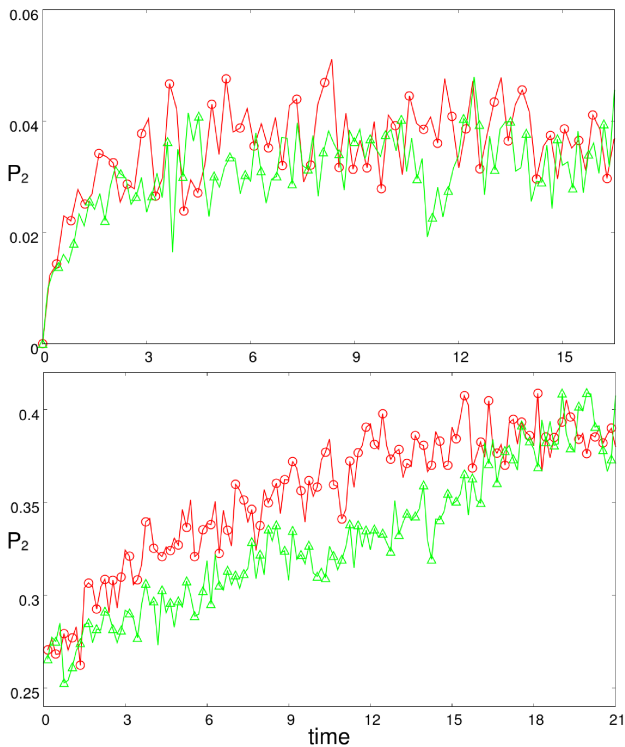}
\caption{\label{fig5} Comparison between BD (circles) and SRD-MD (triangles) results. Upper panel: behavior of $P_2$ in the initial stages of the aggregation process, in which a liquidlike aggregate is formed starting from a diluted randomly dispersed configuration. Lower panel: behavior of $P_2$ for the growth of the NaCl crystal from a liquid-like aggregate in which a well-developed NaCl seed is already present. Time is measured in units of $t^*$.}
\end{figure}

The SRD-MD is much heavier computationally than BD  so that it is very cumbersome to simulate the entire process of two-step aggregation and reorganization, because the nucleation of the ordered phase from the liquid can occur after very long times. For this reason we have decided to restrict the comparison to the analysis of the crucial parts of the process.
First of all, we have started both BD and SRD-MD from initial randomly dispersed non-overlapping configurations and analyzed the initial aggregation stage, leading to the formation of the disordered liquid-like phase, to check whether this phase forms on the same time scale and presents the same properties in both BD and SRD-MD. 
Then we have analyzed the formation of the ordered crystal, starting from configurations in which a NaCl seed is already well developed. These configurations have been taken from previous BD simulations. The configurations with NaCl seed have been let evolve by BD and by SRD-MD.
The results for $\kappa a =3$, averaged over 4 simulations in each case, are reported in Fig. \ref{fig5}, where the behavior of $P_2$ versus time is shown. The figure shows that both BD and SRD-MD produce the same kind of behavior. The liquid-like disordered phase is formed on comparable time scales in both cases, and it presents the same structure, with $P_2$ oscillating around 0.035-0.04. Also the growth of the NaCl seeds in the liquid occurs on the same time scale (in SRD-MD it seems only somewhat slower than in BD), producing the same kind of final structures. These results indicate that hydrodynamic effects do not play a major role in the two-step formation process of the ordered NaCl crystallites.

\section{Conclusions}

In this work we have analyzed the heteroaggregation of oppositely charged colloids by means of a combination of different simulation techniques, including Brownian Dynamics, global optimization searches and Stochastic Rotation Dynamics - Molecular Dynamics.
Aggregation in dilute systems has been considered.
Our results show that colloidal aggregation can be dominated by kinetic effects which can lead to the formation of very well ordered metastable crystals. These crystals significantly differ from the disordered metastable fcc phases that were previously found  in Ref. \cite{Sanz2007prl} at higher volume fractions. 

Our simulations show that, if the range of the potential is sufficiently long ($\kappa a \leq 3.3$), kinetic effects cause the formation of crystallites of the metastable NaCl-type bulk phase, which thus can be obtained also for binary colloids of same size and balanced charge. These crystallites form within a two-step mechanism, in which the first step is the aggregation of a dense metastable liquid whose short-range order has prevailing NaCl features. This triggers the transition to ordered NaCl-type crystallites. Therefore the most likely outcome of the aggregation process is not the crystal structure of greater thermodynamic stability, but the structure whose short-range order is closer to the short-range order in the metastable liquid. This indicates a mechanism of general character which is likely to be relevant for a variety of nucleation processes. The analysis of the most stable structures of small clusters has also shown that the formation of the NaCl phase is not related to 
the structures of very small aggregates, that are either amorphous or of CsCl structure. Our results show also that NaCl-type crystallites more easily form when the NaCl bulk phase is metastable or only marginally stable. Finally, the inclusion of hydrodynamic effects in the simulations has not produced major effects on the aggregation process.
All these findings are important for designing strategies to obtain the desired structures of colloidal crystals in practical cases. 

D. B. acknowledges support 
from the Programme VINCI 2011 of the French-Italian University
on the subject ``Studio dell'eteroaggregazione di particelle colloidali ceramiche tramite tecniche di ottimizzazione globale''.




\begin{thebibliography}{26}
\expandafter\ifx\csname natexlab\endcsname\relax\def\natexlab#1{#1}\fi
\expandafter\ifx\csname bibnamefont\endcsname\relax
  \def\bibnamefont#1{#1}\fi
\expandafter\ifx\csname bibfnamefont\endcsname\relax
  \def\bibfnamefont#1{#1}\fi
\expandafter\ifx\csname citenamefont\endcsname\relax
  \def\citenamefont#1{#1}\fi
\expandafter\ifx\csname url\endcsname\relax
  \def\url#1{\texttt{#1}}\fi
\expandafter\ifx\csname urlprefix\endcsname\relax\def\urlprefix{URL }\fi
\providecommand{\bibinfo}[2]{#2}
\providecommand{\eprint}[2][]{\url{#2}}

\bibitem[{\citenamefont{Leunissen et~al.}(2005)\citenamefont{Leunissen,
  Christova, Hynninen, Royall, Campbell, Imhof, Dijkstra, van Roij, and van
  Blaaderen}}]{Leunissen2005nature}
\bibinfo{author}{\bibfnamefont{M.~E.} \bibnamefont{Leunissen}},
  \bibinfo{author}{\bibfnamefont{C.~G.} \bibnamefont{Christova}},
  \bibinfo{author}{\bibfnamefont{A.~P.} \bibnamefont{Hynninen}},
  \bibinfo{author}{\bibfnamefont{C.~P.} \bibnamefont{Royall}},
  \bibinfo{author}{\bibfnamefont{A.~I.} \bibnamefont{Campbell}},
  \bibinfo{author}{\bibfnamefont{A.}~\bibnamefont{Imhof}},
  \bibinfo{author}{\bibfnamefont{M.}~\bibnamefont{Dijkstra}},
  \bibinfo{author}{\bibfnamefont{R.}~\bibnamefont{van Roij}}, \bibnamefont{and}
  \bibinfo{author}{\bibfnamefont{A.}~\bibnamefont{van Blaaderen}},
  \bibinfo{journal}{Nature} \textbf{\bibinfo{volume}{437}},
  \bibinfo{pages}{235} (\bibinfo{year}{2005}).

\bibitem[{\citenamefont{L\'opez-L\'opez
  et~al.}(2006)\citenamefont{L\'opez-L\'opez, Schmitt, Moncho-Jord\'a, and
  Hidalgo-\'Alvarez}}]{Lopez2006softmatter}
\bibinfo{author}{\bibfnamefont{J.~M.} \bibnamefont{L\'opez-L\'opez}},
  \bibinfo{author}{\bibfnamefont{A.}~\bibnamefont{Schmitt}},
  \bibinfo{author}{\bibfnamefont{A.}~\bibnamefont{Moncho-Jord\'a}},
  \bibnamefont{and}
  \bibinfo{author}{\bibfnamefont{R.}~\bibnamefont{Hidalgo-\'Alvarez}},
  \bibinfo{journal}{Soft Matter} \textbf{\bibinfo{volume}{2}},
  \bibinfo{pages}{1025} (\bibinfo{year}{2006}).

\bibitem[{\citenamefont{Rollie and Sundmacher}(2008)}]{Rollie2008langmuir}
\bibinfo{author}{\bibfnamefont{S.}~\bibnamefont{Rollie}} \bibnamefont{and}
  \bibinfo{author}{\bibfnamefont{K.}~\bibnamefont{Sundmacher}},
  \bibinfo{journal}{Langmuir} \textbf{\bibinfo{volume}{24}},
  \bibinfo{pages}{13348} (\bibinfo{year}{2008}).

\bibitem[{Leb()}]{Lebovka2012note}
\bibinfo{note}{N. I. Lebovka, Adv. Polym. Sci. (2012), DOI:
  10.1007/12\_2012\_171}.

\bibitem[{\citenamefont{Bartlett and Campbell}(2005)}]{Bartlett2005prl}
\bibinfo{author}{\bibfnamefont{P.}~\bibnamefont{Bartlett}} \bibnamefont{and}
  \bibinfo{author}{\bibfnamefont{A.~I.} \bibnamefont{Campbell}},
  \bibinfo{journal}{Phys. Rev. Lett.} \textbf{\bibinfo{volume}{95}},
  \bibinfo{pages}{128302} (\bibinfo{year}{2005}).

\bibitem[{\citenamefont{Hynninen et~al.}(2006)\citenamefont{Hynninen,
  Leunissen, van Blaaderen, and Dijkstra}}]{Hynninen2006prl}
\bibinfo{author}{\bibfnamefont{A.-P.} \bibnamefont{Hynninen}},
  \bibinfo{author}{\bibfnamefont{M.~E.} \bibnamefont{Leunissen}},
  \bibinfo{author}{\bibfnamefont{A.}~\bibnamefont{van Blaaderen}},
  \bibnamefont{and} \bibinfo{author}{\bibfnamefont{M.}~\bibnamefont{Dijkstra}},
  \bibinfo{journal}{Phys. Rev. Lett.} \textbf{\bibinfo{volume}{96}},
  \bibinfo{pages}{018303} (\bibinfo{year}{2006}).

\bibitem[{\citenamefont{Sanz et~al.}(2008)\citenamefont{Sanz, Valeriani,
  Vissers, Fortini, Leunissen, van Blaaderen, Frenkel, and
  Dijkstra}}]{Sanz2008jpcm}
\bibinfo{author}{\bibfnamefont{E.}~\bibnamefont{Sanz}},
  \bibinfo{author}{\bibfnamefont{C.}~\bibnamefont{Valeriani}},
  \bibinfo{author}{\bibfnamefont{T.}~\bibnamefont{Vissers}},
  \bibinfo{author}{\bibfnamefont{A.}~\bibnamefont{Fortini}},
  \bibinfo{author}{\bibfnamefont{M.~E.} \bibnamefont{Leunissen}},
  \bibinfo{author}{\bibfnamefont{A.}~\bibnamefont{van Blaaderen}},
  \bibinfo{author}{\bibfnamefont{D.}~\bibnamefont{Frenkel}}, \bibnamefont{and}
  \bibinfo{author}{\bibfnamefont{M.}~\bibnamefont{Dijkstra}},
  \bibinfo{journal}{J. Phys.: Condens. Matter} \textbf{\bibinfo{volume}{20}},
  \bibinfo{pages}{494247} (\bibinfo{year}{2008}).

\bibitem[{\citenamefont{Piechowiak et~al.}(2012)\citenamefont{Piechowiak,
  Videcoq, Ferrando, Bochicchio, Pagnoux, and Rossignol}}]{Piechowiak2012pccp}
\bibinfo{author}{\bibfnamefont{M.~A.} \bibnamefont{Piechowiak}},
  \bibinfo{author}{\bibfnamefont{A.}~\bibnamefont{Videcoq}},
  \bibinfo{author}{\bibfnamefont{R.}~\bibnamefont{Ferrando}},
  \bibinfo{author}{\bibfnamefont{D.}~\bibnamefont{Bochicchio}},
  \bibinfo{author}{\bibfnamefont{C.}~\bibnamefont{Pagnoux}}, \bibnamefont{and}
  \bibinfo{author}{\bibfnamefont{F.}~\bibnamefont{Rossignol}},
  \bibinfo{journal}{Phys. Chem. Chem. Phys.} \textbf{\bibinfo{volume}{14}},
  \bibinfo{pages}{1431} (\bibinfo{year}{2012}).

\bibitem[{\citenamefont{Russell et~al.}(2012)\citenamefont{Russell, Sprakel,
  Kodger, and Weitz}}]{Russell2012softmatter}
\bibinfo{author}{\bibfnamefont{E.~R.} \bibnamefont{Russell}},
  \bibinfo{author}{\bibfnamefont{J.}~\bibnamefont{Sprakel}},
  \bibinfo{author}{\bibfnamefont{T.~E.} \bibnamefont{Kodger}},
  \bibnamefont{and} \bibinfo{author}{\bibfnamefont{D.~A.} \bibnamefont{Weitz}},
  \bibinfo{journal}{Soft Matter} \textbf{\bibinfo{volume}{8}},
  \bibinfo{pages}{8697} (\bibinfo{year}{2012}).

\bibitem[{\citenamefont{Bier et~al.}(2010)\citenamefont{Bier, van Roij, and
  Dijkstra}}]{Bier2010jcp}
\bibinfo{author}{\bibfnamefont{M.}~\bibnamefont{Bier}},
  \bibinfo{author}{\bibfnamefont{R.}~\bibnamefont{van Roij}}, \bibnamefont{and}
  \bibinfo{author}{\bibfnamefont{M.}~\bibnamefont{Dijkstra}},
  \bibinfo{journal}{J. Chem. Phys.} \textbf{\bibinfo{volume}{133}},
  \bibinfo{pages}{124501} (\bibinfo{year}{2010}).

\bibitem[{\citenamefont{Pavaskar et~al.}(2012)\citenamefont{Pavaskar, Sharma,
  and Punnathanam}}]{Pavaskar2012jcp}
\bibinfo{author}{\bibfnamefont{G.}~\bibnamefont{Pavaskar}},
  \bibinfo{author}{\bibfnamefont{S.}~\bibnamefont{Sharma}}, \bibnamefont{and}
  \bibinfo{author}{\bibfnamefont{S.~N.} \bibnamefont{Punnathanam}},
  \bibinfo{journal}{J. Chem. Phys.} \textbf{\bibinfo{volume}{136}},
  \bibinfo{pages}{134506} (\bibinfo{year}{2012}).

\bibitem[{\citenamefont{Sanz et~al.}(2007)\citenamefont{Sanz, Valeriani,
  Frenkel, , and Dijkstra}}]{Sanz2007prl}
\bibinfo{author}{\bibfnamefont{E.}~\bibnamefont{Sanz}},
  \bibinfo{author}{\bibfnamefont{C.}~\bibnamefont{Valeriani}},
  \bibinfo{author}{\bibfnamefont{D.}~\bibnamefont{Frenkel}}, ,
  \bibnamefont{and} \bibinfo{author}{\bibfnamefont{M.}~\bibnamefont{Dijkstra}},
  \bibinfo{journal}{Phys. Rev. Lett.} \textbf{\bibinfo{volume}{99}},
  \bibinfo{pages}{055501} (\bibinfo{year}{2007}).

\bibitem[{\citenamefont{Cerbelaud
  et~al.}(2010{\natexlab{a}})\citenamefont{Cerbelaud, Videcoq, Ab\'elard,
  Pagnoux, Rossignol, and Ferrando}}]{Cerbelaud2010softmatter}
\bibinfo{author}{\bibfnamefont{M.}~\bibnamefont{Cerbelaud}},
  \bibinfo{author}{\bibfnamefont{A.}~\bibnamefont{Videcoq}},
  \bibinfo{author}{\bibfnamefont{P.}~\bibnamefont{Ab\'elard}},
  \bibinfo{author}{\bibfnamefont{C.}~\bibnamefont{Pagnoux}},
  \bibinfo{author}{\bibfnamefont{F.}~\bibnamefont{Rossignol}},
  \bibnamefont{and} \bibinfo{author}{\bibfnamefont{R.}~\bibnamefont{Ferrando}},
  \bibinfo{journal}{Soft Matter} \textbf{\bibinfo{volume}{6}},
  \bibinfo{pages}{370} (\bibinfo{year}{2010}{\natexlab{a}}).

\bibitem[{\citenamefont{Piechowiak et~al.}(2010)\citenamefont{Piechowiak,
  Videcoq, Rossignol, Pagnoux, , Carrion, Cerbelaud, and
  Ferrando}}]{Piechowiak2010langmuir}
\bibinfo{author}{\bibfnamefont{M.~A.} \bibnamefont{Piechowiak}},
  \bibinfo{author}{\bibfnamefont{A.}~\bibnamefont{Videcoq}},
  \bibinfo{author}{\bibfnamefont{F.}~\bibnamefont{Rossignol}},
  \bibinfo{author}{\bibfnamefont{C.}~\bibnamefont{Pagnoux}}, ,
  \bibinfo{author}{\bibfnamefont{C.}~\bibnamefont{Carrion}},
  \bibinfo{author}{\bibfnamefont{M.}~\bibnamefont{Cerbelaud}},
  \bibnamefont{and} \bibinfo{author}{\bibfnamefont{R.}~\bibnamefont{Ferrando}},
  \bibinfo{journal}{Langmuir} \textbf{\bibinfo{volume}{26}},
  \bibinfo{pages}{12540} (\bibinfo{year}{2010}).

\bibitem[{\citenamefont{Allen et~al.}(2005)\citenamefont{Allen, Warren, and ten
  Wolde}}]{Allen2005prl}
\bibinfo{author}{\bibfnamefont{R.~J.} \bibnamefont{Allen}},
  \bibinfo{author}{\bibfnamefont{P.~B.} \bibnamefont{Warren}},
  \bibnamefont{and} \bibinfo{author}{\bibfnamefont{P.~R.} \bibnamefont{ten
  Wolde}}, \bibinfo{journal}{Phys. Rev. Lett.} \textbf{\bibinfo{volume}{94}},
  \bibinfo{pages}{018104} (\bibinfo{year}{2005}).

\bibitem[{\citenamefont{Lutsko and Nicolis}(2006)}]{Lutsko2006prl}
\bibinfo{author}{\bibfnamefont{J.~F.} \bibnamefont{Lutsko}} \bibnamefont{and}
  \bibinfo{author}{\bibfnamefont{G.}~\bibnamefont{Nicolis}},
  \bibinfo{journal}{Phys. Rev. Lett.} \textbf{\bibinfo{volume}{96}},
  \bibinfo{pages}{046102} (\bibinfo{year}{2006}).

\bibitem[{\citenamefont{Vekilov}(2010)}]{Vekilov2010nanoscale}
\bibinfo{author}{\bibfnamefont{P.~G.} \bibnamefont{Vekilov}},
  \bibinfo{journal}{Nanoscale} \textbf{\bibinfo{volume}{2}},
  \bibinfo{pages}{2346} (\bibinfo{year}{2010}).

\bibitem[{\citenamefont{Schilling et~al.}(2010)\citenamefont{Schilling,
  Sch\"ope, Oettel, Opletal, and Snook}}]{Schilling2010prl}
\bibinfo{author}{\bibfnamefont{T.}~\bibnamefont{Schilling}},
  \bibinfo{author}{\bibfnamefont{H.~J.} \bibnamefont{Sch\"ope}},
  \bibinfo{author}{\bibfnamefont{M.}~\bibnamefont{Oettel}},
  \bibinfo{author}{\bibfnamefont{G.}~\bibnamefont{Opletal}}, \bibnamefont{and}
  \bibinfo{author}{\bibfnamefont{I.}~\bibnamefont{Snook}},
  \bibinfo{journal}{Phys. Rev. Lett.} \textbf{\bibinfo{volume}{105}},
  \bibinfo{pages}{025701} (\bibinfo{year}{2010}).

\bibitem[{\citenamefont{Vermolen et~al.}(2009)\citenamefont{Vermolen, Kuijk,
  Filion, Hermes, Thijssen, Dijkstra, and van Blaaderen}}]{Vermolen2009pnas}
\bibinfo{author}{\bibfnamefont{E.~C.~M.} \bibnamefont{Vermolen}},
  \bibinfo{author}{\bibfnamefont{A.}~\bibnamefont{Kuijk}},
  \bibinfo{author}{\bibfnamefont{L.~C.} \bibnamefont{Filion}},
  \bibinfo{author}{\bibfnamefont{M.}~\bibnamefont{Hermes}},
  \bibinfo{author}{\bibfnamefont{J.~H.~J.} \bibnamefont{Thijssen}},
  \bibinfo{author}{\bibfnamefont{M.}~\bibnamefont{Dijkstra}}, \bibnamefont{and}
  \bibinfo{author}{\bibfnamefont{A.}~\bibnamefont{van Blaaderen}},
  \bibinfo{journal}{Proc. Natl. Acad. Sci.} \textbf{\bibinfo{volume}{106}},
  \bibinfo{pages}{16063} (\bibinfo{year}{2009}).

\bibitem[{\citenamefont{Malevanets and Kapral}(2000)}]{Sh:1}
\bibinfo{author}{\bibfnamefont{A.}~\bibnamefont{Malevanets}} \bibnamefont{and}
  \bibinfo{author}{\bibfnamefont{J.}~\bibnamefont{Kapral}},
  \bibinfo{journal}{J. Chem. Phys.} \textbf{\bibinfo{volume}{112}},
  \bibinfo{pages}{7260} (\bibinfo{year}{2000}).

\bibitem[{\citenamefont{Padding and Louis}(2006)}]{Sh:2}
\bibinfo{author}{\bibfnamefont{J.~T.} \bibnamefont{Padding}} \bibnamefont{and}
  \bibinfo{author}{\bibfnamefont{A.~A.} \bibnamefont{Louis}},
  \bibinfo{journal}{Phys. Rev. E} \textbf{\bibinfo{volume}{74}},
  \bibinfo{pages}{031402} (\bibinfo{year}{2006}).

\bibitem[{\citenamefont{Tomilov et~al.}(2012)\citenamefont{Tomilov, Videcoq,
  Chartier, Ala-Nissila, and Vattulainen}}]{Sh:3}
\bibinfo{author}{\bibfnamefont{A.}~\bibnamefont{Tomilov}},
  \bibinfo{author}{\bibfnamefont{A.}~\bibnamefont{Videcoq}},
  \bibinfo{author}{\bibfnamefont{T.}~\bibnamefont{Chartier}},
  \bibinfo{author}{\bibfnamefont{T.}~\bibnamefont{Ala-Nissila}},
  \bibnamefont{and}
  \bibinfo{author}{\bibfnamefont{I.}~\bibnamefont{Vattulainen}},
  \bibinfo{journal}{J Chem. Phys.} \textbf{\bibinfo{volume}{137}},
  \bibinfo{pages}{014503} (\bibinfo{year}{2012}).

\bibitem[{\citenamefont{Wales and Doye}(1997)}]{Wales1997}
\bibinfo{author}{\bibfnamefont{D.~J.} \bibnamefont{Wales}} \bibnamefont{and}
  \bibinfo{author}{\bibfnamefont{J.~P.~K.} \bibnamefont{Doye}},
  \bibinfo{journal}{J.Phys. Chem. A} \textbf{\bibinfo{volume}{101}},
  \bibinfo{pages}{5111} (\bibinfo{year}{1997}).

\bibitem[{\citenamefont{Cerbelaud
  et~al.}(2010{\natexlab{b}})\citenamefont{Cerbelaud, Ferrando, and
  Videcoq}}]{Cerbelaud2010jcp}
\bibinfo{author}{\bibfnamefont{M.}~\bibnamefont{Cerbelaud}},
  \bibinfo{author}{\bibfnamefont{R.}~\bibnamefont{Ferrando}}, \bibnamefont{and}
  \bibinfo{author}{\bibfnamefont{A.}~\bibnamefont{Videcoq}},
  \bibinfo{journal}{J. Chem. Phys.} \textbf{\bibinfo{volume}{132}},
  \bibinfo{pages}{084701} (\bibinfo{year}{2010}{\natexlab{b}}).

\bibitem[{\citenamefont{Rossi and Ferrando}(2009)}]{Rossi09jpcm}
\bibinfo{author}{\bibfnamefont{G.}~\bibnamefont{Rossi}} \bibnamefont{and}
  \bibinfo{author}{\bibfnamefont{R.}~\bibnamefont{Ferrando}},
  \bibinfo{journal}{J. Phys. Cond. Mat.} \textbf{\bibinfo{volume}{21}},
  \bibinfo{pages}{084208} (\bibinfo{year}{2009}).

\bibitem[{\citenamefont{Sanz et~al.}(2011)\citenamefont{Sanz, Valeriani,
  Zaccarelli, Poon, Pusey, and Cates}}]{Sanz2011prl}
\bibinfo{author}{\bibfnamefont{E.}~\bibnamefont{Sanz}},
  \bibinfo{author}{\bibfnamefont{C.}~\bibnamefont{Valeriani}},
  \bibinfo{author}{\bibfnamefont{E.}~\bibnamefont{Zaccarelli}},
  \bibinfo{author}{\bibfnamefont{W.~C.~K.} \bibnamefont{Poon}},
  \bibinfo{author}{\bibfnamefont{P.~N.} \bibnamefont{Pusey}}, \bibnamefont{and}
  \bibinfo{author}{\bibfnamefont{M.~E.} \bibnamefont{Cates}},
  \bibinfo{journal}{Phys. Rev. Lett.} \textbf{\bibinfo{volume}{106}},
  \bibinfo{pages}{215701} (\bibinfo{year}{2011}).

\end{thebibliography}

\end{document}